\documentclass[aps,twocolumn,10pt,aps,amsmath,amssymb,nofootinbib,superscriptaddress,showkeys,showpacs]{revtex4-1}
\usepackage{epsfig} 
\usepackage{amsmath} 
\usepackage{graphicx}
\usepackage{color}
\usepackage{float}
\usepackage{soul,xcolor}
\usepackage[font=scriptsize]{caption}
\usepackage{braket}
\usepackage{bbold}
\usepackage{subfig}
\usepackage{braket}
\usepackage{titlesec}
\usepackage{placeins}
\usepackage{caption}
\usepackage[normalem]{ulem}
\begin{document} 
	
	\title{New physics effects on quantum coherence in neutrino oscillations}

	\author{Khushboo Dixit}
	\email{dixit.1@iitj.ac.in}
	\affiliation{Indian Institute of Technology Jodhpur, Jodhpur 342037, India}
	
	\author{Ashutosh Kumar Alok}
	\email{akalok@iitj.ac.in}
	\affiliation{Indian Institute of Technology Jodhpur, Jodhpur 342037, India}
	
	\date{\today}

	\begin{abstract}
		Several measures of quantum correlations such as Leggett-Garg and Bell-type inequalities have been extensively studied in the context of neutrino oscillations. However these analyses are performed under the assumption of standard model (SM) interactions of neutrinos. In this work we study new physics effects on $l_1$-norm based measure of quantum coherence which quantifies the quantumness embedded in the system and is also intrinsically related to various measures of quantum correlations. Moreover, it is considered to be a resource theoretical tool which can be utilized in quantum algorithms and quantum channel discrimination. The new physics effects are incorporated in a model independent way by using the effective Lagrangian for the neutral current non-standard neutrino interactions (NSI). Bounds on the NSI parameters are extracted from a recent global analysis of oscillation experiments including COHERENT (coherent neutrino-nucleus scattering experiment) data. In the context of upcoming DUNE experimental setup, we find that the most favourable combination of LMA-Light sector of $\theta_{12}$ ({\it i.e.}, $\theta_{12}< 45^o$) with normal mass ordering decreases the coherence in the system in comparison to the SM prediction for all values of neutrino energy $E$ and $CP$ violating phase $\delta$ (except in the narrow region around $E \sim$ 2 GeV). On the other hand, a large enhancement in the value of coherence parameter in the entire $(E-\delta)$ plane is possible for the dark octant of $\theta_{12}$ ($\theta_{12}> 45^o$) with inverted ordering. For almost all values of $CP$ violating phase, the enhancement is more protuberant in the region around $E \sim$ 4 GeV where maximum neutrino flux is expected in the DUNE experiment. Therefore for the normal mass ordering, the SM interaction provides favourable conditions for quantum information tasks while the NSI favours inverted ordering scenario for such tasks.
	\end{abstract}
	
	\maketitle

	\section{Introduction}
	
The phenomena of neutrino oscillation is a consequence of superposition principle which makes the quantum coherence an indispensable part of the system. It implies that neutrinos can change their flavour after traveling a macroscopically large distance which is attributed to the nonzero neutrino mass and neutrino mixing. Till date, evidences for this phenomenon have been collected in various solar \cite{Ahmad}, atmospheric \cite{Fukuda}, reactor \cite{Araki} and long baseline accelerator experiments \cite{Abe}. Also, these experimental facilities have measured the oscillation parameters with a good accuracy and  we are now tending towards more precise measurements. Moreover, due to weakly interacting nature, the system of oscillating neutrinos can maintain quantum coherence over a long distance \cite{Kaiser} and hence can turn out to be promising and better candidates for various tasks related to quantum information\footnote{Although the dynamics of the mesonic system is also driven by weak interactions, owing to its short lifetime, this system would be more suitable for understanding foundational issues rather than having any applicational implications.} compared to the photonic or electronic systems. The transmission of information using NuMI beam and MINER$\nu$A detector at Fermilab has been demonstrated by Stancil et al. in \cite{Stancil}. To explore the possibilities of the utilization of neutrino system to achieve future endeavours in quantum information sector, it is pertinent to evaluate the quantumness of the system.

On this account, several measures of quantum correlations such as entanglement \cite{Horodecki}, concurrence \cite{Wootters}, Leggett-Garg inequality \cite{Leggett} and Bell's inequality \cite{Bell} have been constructed in order to test the quantumness of the system. These measures have been studied previously in the context of neutrino system \cite{Blasone1,Blasone2,Alok,Banerjee,Dixit2,Kaiser,dayabay,Naikoo1,Richter-Laskowska:2018ikv,Blasone:2018ktu,Naikoo2}. The experimentally observed neutrino oscillations have shown violation of classical bounds of Leggett-Garg and Bell-type inequalities in these studies. However, the degree of violation of these inequalities cannot be considered as a measure to quantify the quantummess of the system because in certain cases their maximum value depends on the channel parameters \cite{Cereceda,Emary,Song}, {\it i.e.}, these quantities can only witness the quantum nature of a given system rather than quantifying it.

In the context of quantum information, coherence is a fundamental concept which can quantify the quantumness of the system. It can be rigorously characterized in the context of quantum resource theory. Also, quantum coherence is closely related to various measures of quantum correlations, such as entanglement and quantum discord \cite{Streltsov}. Expressly, quantum coherence can be considered as linchpin of all the quantum correlations shared within the system. In the context of quantum information theory, the quantification of coherence can be done based on the characterization of the set of incoherent states ($\mathcal{I}$) and incoherent operations ($\mathcal{O^I}$). In a given reference basis $\{\ket{i}\}$, states defined as $\rho_{\mathcal{I}} = \sum_{i} d_i\ket{i}\bra{i}$ where $d_i \geq 0$ and $\sum_{i} d_i = 1$, form a set of incoherent states. Incoherent operations are specified in such a manner that they map the set of incoherent states onto itself, {\it i.e.}, $\mathcal{O^I}(\mathcal{I}) \in \mathcal{I}$. In such a set of incoherent operations ($\mathcal{O^I}$) and incoherent states ($\mathcal{I}$), the $l_1$-norm (sum of the absolute values of off-diagonal elements of the density matrix $\rho=\ket{\psi} \bra{\psi}$) \cite{Baumgratz,Uttam}
\begin{equation}
	\chi = \sum_{i\neq j}|\rho_{ij}| \geq 0,
	\label{coherence}
\end{equation}
represents a reliable measure or quantifier of coherence. It also acquires all the basic properties of a coherence-measure such as, (i) non-negativity, {\it i.e.} $\chi(\rho) = 0$ iff $\rho \in \mathcal{I}$, (ii) monotonicity under incoherent operations, in other words, $\chi(\rho)$ is nonincreasing under the incoherent operations, {\it i.e.}, $\chi(\mathcal{O^I}[\rho]) \leq \chi(\rho)$ and (iii) convexity, {\it i.e.}, $\chi(\sum_{k}p_k\rho_k) \leq \sum_{k}p_k\chi(\rho_k)$, where $\rho_k = Q_k \rho Q_k^{\dagger}/p_k$ ($Q_k$ is the Kraus operator) and $p_k = Tr(Q_k \rho Q_k^{\dagger})$. The value of this measure for a d-dimensional maximally coherent state, defined by $\ket{\psi_d} = \frac{1}{\sqrt{d}}\sum_{i=1}^{d}\ket{i}$, becomes $d-1$ placing an upper bound (or maximal value) of this measure. While $\chi$ is zero for a completely incoherent state. For three flavour neutrino oscillation scenario, $\chi^{max} = 2$. 

Recently, the $l_1$-norm based measure of quantum coherence has been quantified in terms of experimentally observed neutrino survival and transition probabilities \cite{Song}. Furthermore, quantum coherence is considered to be a resource theoretical tool which can be utilized in quantum algorithms \cite{Matera} and quantum channel discrimination \cite{Piani}. Similar to quantum entanglement distillation, one can estimate the number of copies to transform the less coherent states into maximally coherent ones as needed through incoherent operations \cite{Winter}. In this case, it becomes important to quantify the quantum coherence of the system. Also, it has been seen that the quantum information theoretic quantities can help to resolve the open problems in neutrino sector such as mass hierarchy problem \cite{Naikoo1,Dixit2} and discrimination between Dirac and Majorana nature of neutrinos \cite{Richter}.

However, so far, the studies related to quantum correlations in neutrino physics are performed assuming that the dynamics of the system is governed only by the Standard Model (SM) physics. The fact that SM cannot be considered as the quintessential theory of fundamental interactions of nature, one needs to explore the effects of beyond SM physics.
The SM Lagrangian contains only renormalizable interactions with canonical dimensions $D \leq 4$. Assuming that new physics (NP) exists at some high energy $\Lambda$, the effects of these NP interactions at the energy scale much below $\Lambda$, can be described in a model independent way by including higher dimensional operators constructed out of the SM fields. The first observable consequence of such NP emerges in terms of nonzero neutrino masses after electroweak symmetry breaking in the form of the Weinberg operator \cite{Weinberg:1979sa}, the only dimension-5 operator composed of SM fields. However, to generate a small neutrino mass $\sim 1$ eV, the required NP scale is $10^{13}$ GeV which is far beyond the energy scale achieved in LHC till date \cite{Buchmuller:1985jz,Krauss:2013lra,Babu:2009aq}. Next higher dimensional operator with visible effects for energy $<<\Lambda$ appears as non-renormalizable lepton number conserving four-fermion-dimension-6 operators leading to nonstandard neutrino interactions (NSI) with matter \cite{Wolfenstein,Valle,Guzzo,Grossman:1995wx,Krastev,Brooijmans,Gonzalez,Bergmann,Guzzo2,Guzzo3,Antusch:2008tz,Ohlsson:2012kf,Miranda:2015dra,Farzan:2015doa,Farzan:2015hkd,Babu:2017olk,Denton:2018xmq,Farzan:2017xzy,Falkowski:2018dmy,Esteban,Esteban3}.

A global analysis of oscillation data with neutral current NSI interactions in three flavour scenario was performed in \cite{Esteban} where observables sensitive to the $CP$-violating phase (such as $\nu_e$ and $\bar{\nu}_e$ appearance at long baseline experiments) were excluded from the fit and hence the constraints were obtained on the $CP$-conserving part only. Also, some approximations ($\Delta_{21}  = m_2^2 - m_1^2 \rightarrow 0$ in atmospheric and long baseline $CP$-conserving experiments, where $m_1$ and $m_2$ are eigenvalues corresponding to the neutrino mass eigenstates $\ket{\nu_1}$ and $\ket{\nu_2}$) were used to simplify the calculations, hence the effect of mass ordering was greatly reduced. This analysis was performed for both scenarios: first and second octant solution of solar mixing angle $\theta_{12}$. Recently this global analysis has been extended to include complex NSI neutral current interaction with quarks (interactions with electrons were excluded in both of these analyses) for observables sensitive to the leptonic $CP$-violating phase and mass ordering \cite{Esteban3}. In \cite{Esteban3} they analyzed all the four combinations, {\it i.e.}, light ($\theta_{12}\le 45^o$) and dark ($\theta_{12}\ge 45^o$) sector with normal ordering ($m_1<m_2<m_3$) (NO) and inverted ordering ($m_3<m_1<m_2$) (IO) of mass states, in which two solutions, light octant with NO and dark octant with IO, are favoured in global analysis of oscillation data\footnote{Solar neutrino data is found to disfavor the LMA-Dark solution with confidence level below 2$\sigma$. However, LMA-D provides an equally good fit to the LBL data (specifically T2K data). Hence we have considered LMA-Light with NO and LMA-Dark with IO as most favoured solutions.}.  

The fact that the experimental facilities in neutrino physics are now tending towards higher precision and have potential to probe sub leading effects like non-standard neutrino-matter interaction, it is worth considering NP effects on various measures of quantum correlations in the context of neutrino systems. Testing the quantum coherence to analyze NP effects on the quantumness sustained in the system can provide an idea of overall behavior of various quantum correlations in the neutrino-system under the influence of such effects. In this work we study the effects of NSI on quantum coherence embedded in the neutrino system quantified in terms of $l_1$-norm of coherence. Moreover, the methodology of this work can serve as a guideline for implementing NP effects in the study of various other measures of quantum correlations in neutrinos such as non-locality, entanglement and discord. In view of recent updates, we study the effects of neutral current NSI, relevant for neutrino propagation in matter, on the quantum coherence of the neutrino system within the context of long baseline Deep Underground Neutrino Experiment (DUNE) \cite{Acciarri:2015uup} and identify scenarios ({\it i.e.}, ranges of neutrino-energy and $CP$-phase $\delta$) where coherence is maximal.

The plan of the paper is as follows. We start with the general formalism to incorporate the NSI effects in the dynamics of neutrino oscillations in Sec. \ref{NueOsc}. Then in Sec. \ref{MI}, $l_1$-norm of coherence is calculated for the neutrino system with NSI effects. Further, we present the analysis of NSI effects on this coherence parameter in the context of DUNE experimental set-up and discuss our results. Finally, we conclude in Sec. \ref{Conclusion}.

\section{NSI effects on neutrino oscillations}\label{NueOsc}
	
	The NP neutrino-matter interactions can occur due to charged current (CC) as well as neutral current (NC) interactions. Both NSI-NC and NSI-CC can modify the inelastic neutrino scattering cross sections with other SM fermions \cite{Davidson:2003ha,Biggio:2009nt,Biggio:2009kv}. However, the mean-free path for inelastic interactions of neutrino in earth's matter is much larger than the earth's diameter for neutrinos with energy $< 10^5$ GeV \cite{Giunti:2007ry}. Hence, effects of such scatterings can be neglected. The NSI-NC with two neutrinos can also affect the forward coherent scattering as neutrinos propagate through matter via so called Mikheev-Smirnov-Wolfenstein (MSW) mechanism \cite{Wolfenstein,Valle,Guzzo,Ohlsson:2012kf,Miranda:2015dra,Davidson:2003ha,Biggio:2009nt,Biggio:2009kv,Mikheev,Guzzo1,Roulet}. Consequently, a significantly enhanced effect of NSI-NC can be seen in long baseline oscillation experiments, such as DUNE, where neutrinos have to travel through a large region of matter. The charged-current NSI of neutrinos with matter ({\it i.e.}, e,u,d) can affect the production and detection of neutrinos (in general called zero distance effect) and can become discernible in near detectors. Moreover, the scattering bounds on NSI-CC are rather stringent due to constraints coming from Fermi constant, pion decay, CKM unitarity and the kinematic measurements of the masses of the gauge bosons $M_Z$ and $M_W$, while these bounds are  weaker for NSI-NC approximately by one order of magnitude. The concept of NSI was originated in \cite{Wolfenstein} introducing the {\it flavor changing neutral current}. In this work, we consider the effects of neutral-current interactions driven by NSI relevant to neutrino propagation in matter which can be represented by the Lagrangian
	\begin{equation}
		\mathcal{L}_{NSI} = -2\sqrt{2} G_F \sum_{f,P,\alpha,\beta}\epsilon_{\alpha,\beta}^{f,P} (\bar{\nu}_{\alpha}\gamma^{\mu}P_L \nu_{\beta})(\bar{f}\gamma_{\mu}P f),
	\end{equation}
	where $G_F$ is the Fermi constant, $\alpha$ and $\beta$ are flavour indices, $P_L$ \& $P_R$ are the projection operators and $f$ is the charged fermion. Here, $\epsilon_{\alpha,\beta}^{f,P} \sim \mathcal{O} (G_x/G_F)$ represents the strength of the new interaction with respect to the SM interaction which is quantified by $G_F$. If the flavour of neutrinos participating in the interaction is considered to be independent of the charged fermion type, one can write $\epsilon_{\alpha\beta}^{f,P} \equiv \epsilon_{\alpha\beta}^{\eta}\, \xi^{f,P}$, where matrix elements $\epsilon^{\eta}_{\alpha\beta}$ correspond to the coupling between neutrinos and the coefficients $\xi^{f,P}$ represent the coupling to the charged fermions. Hence the Lagrangian becomes
	\begin{align}
		\mathcal{L}_{NSI} =& -2\sqrt{2} G_F \sum_{\alpha,\beta}\epsilon_{\alpha,\beta}^{\eta} (\bar{\nu}_{\alpha}\gamma^{\mu}P_L \nu_{\beta})
		\\ \nonumber
		&\times \sum_{f,P}\xi^{f,P}(\bar{f}\gamma_{\mu}P f).
	\end{align} 
	
	The Hamiltonian for the evolution of neutrino-state, in mass eigenstate basis, including NSI effect can be written as $\mathcal{H}_m = H_{m} + U_v^{-1} V_f U_v$,  where $H_m = \rm diag(E_1, E_2, E_3)$ and $U_v$ is the 3$\times$3 unitary (PMNS) matrix. The matter part $V_f$ of the Hamiltonian including the operators corresponding to the NSI effect is given as
	\begin{equation}
		V_f = A \begin{pmatrix}
			1+\epsilon_{ee}(x)      & \epsilon_{e\mu}(x)      & \epsilon_{e\tau}(x)\\
			\epsilon_{e\mu}^*(x)  & \epsilon_{\mu\mu}(x)    & \epsilon_{\mu\tau}(x)\\
			\epsilon_{e\tau}^*(x)  & \epsilon_{\mu\tau}^*(x) & \epsilon_{\tau\tau}(x)
		\end{pmatrix},
	\end{equation}
	with $A = \sqrt{2}G_F N_e(x)$. Here, ``1'' in the $1\times1$ element of $V_f$ corresponds to the standard matter interaction of neutrinos and 
	\begin{equation}
		\epsilon_{\alpha \beta} = \sum_{f=e,u,d}\frac{N_f(x)}{N_e(x)}\epsilon_{\alpha\beta}^f,
		\label{NSIparameter}
	\end{equation}
	represents the non-standard part. Here, $N_f(x)$ is the number density of fermion $f$ as a function of the distance $x$ traveled by neutrino. According to the quark-structure of protons (p) and neutrons (n), we can write
	\begin{equation}
		N_u(x) = 2 N_p(x) + N_n(x), ~~~~~~~~ N_d(x) = N_p(x) + 2 N_n(x).
		\label{density}
	\end{equation} 
	Therefore, from Eq. (\ref{NSIparameter}) and (\ref{density}) we can write
	\begin{equation}
		\epsilon_{\alpha\beta} = (2 + Y_n) \epsilon_{\alpha\beta}^u + (1 + 2 Y_n) \epsilon_{\alpha\beta}^d,
	\end{equation}
	with $Y_n = N_n/N_e$, $N_e$ is the number density of electrons and $N_p = N_e$.  Here we used a different parameterization of mixing matrix which differs from the usual one $U$ by an overall phase matrix $P = \rm diag(e^{i\delta},1,1)$ by $U_v = P U P^{\ast}$ and represented as,
	
	\small
	\begin{equation}
		U_v = 
		\begin{pmatrix}
			c_{12}c_{13}                               &s_{12}c_{13}e^{i\delta}                  &s_{13}\\
			-s_{12}c_{23}e^{-i\delta}-c_{12}s_{13}s_{23}    &c_{12}c_{23}-s_{12}s_{13}s_{23}e^{i\delta}       &c_{13}s_{23}\\
			s_{12}s_{23}e^{-i\delta}-c_{12}s_{13}c_{23}     &-c_{12}s_{23}-s_{12}s_{13}c_{23}e^{i\delta}      &c_{13}c_{23}
		\end{pmatrix}.
	\end{equation}
	\normalsize
	
	This rephasing does not affect the probability expressions in the absence of NSI. The advantage of this convention of $U$-matrix is that one can easily perform the $CPT$-transformation, $H_{vac} \rightarrow -H_{vac}^{\ast}$, as just by doing simple replacements of the oscillation parameters (mixing angles $\theta_{ij}$ and mass squared differences $\Delta_{ij} = m_j^2-m_i^2$), such as\\
	\begin{eqnarray}
		\Delta_{31} \rightarrow -\Delta_{31} + \Delta_{21} \rightarrow -\Delta_{32},\nonumber\\ 
		\theta_{12} \rightarrow \pi/2-\theta_{12},~~~~~~~~~~~ \nonumber\\ 
		\delta \rightarrow \pi-\delta.~~~~~~~~~~~~~~~~~~ 
		\label{CPTvac}
	\end{eqnarray}

	The $CPT$-transformation of $H_{vac}$, in which neutrino evolution remains invariant, involves the change of the octant of $\theta_{12}$ (dark octant with $\theta_{12}>45^o$) and also the change in the sign of $\Delta_{31}$. The octant selection of mixing-angle $\theta_{12}$ becomes important when neutrino is traveling through a dense material medium as the possibility of NSI increases. For example, the deficit of solar neutrinos at the detectors can be resolved by considering the vacuum mixing angle in the light-side ($0\leq \theta_{12}\leq \frac{\pi}{4}$) with standard neutrino-matter interactions as well as the dark-side solution ($\frac{\pi}{4}\leq \theta_{12}\leq \frac{\pi}{2}$) with large enough values of NSI parameters \cite{Miranda_2006}. This specific feature is called the {\it generalized mass ordering degeneracy} that was first noticed in \cite{Coloma1,Coloma2}.
	
	To include $CPT$-transformation in matter-part of Hamiltonian, the replacements are 
	\begin{align}
		[\epsilon_{ee} - \epsilon_{\mu\mu}] \rightarrow - [\epsilon_{ee} - \epsilon_{\mu\mu}] - 2, \nonumber\\ 
		[\epsilon_{\tau\tau} - \epsilon_{\mu\mu}] \rightarrow - [\epsilon_{\tau\tau} - \epsilon_{\mu\mu}],~~~~\nonumber\\ 
		\epsilon_{\alpha\beta} \rightarrow - \epsilon_{\alpha\beta}^*   ~(\alpha \neq \beta).~~~~~~~~~~~~
	\end{align}
	The evolution of mass eigenstate $\psi_m$ can be given by
	\begin{equation}
		\psi_m(L) = e^{-i \mathcal{H}_m L} \psi_m(0) \equiv \mathcal{U}_m(L) \psi_m(0).
	\end{equation}
	In order to obtain the evolution operator $\mathcal{U}_m$ in the mass eigenstate basis,  we use the formalism given in \cite{Ohlsson}. Using Cayley-Hamilton's theorem, which implies that, in the characteristic equation of a N$\times$N matrix $M$, {\it i.e.}, ${\rm det}(M-\lambda \textbf{I}) = 0$, the eigenvalue $\lambda$ can be replaced by the matrix $M$ itself, hence reducing the number of terms in the exponential series $e^M$ to N. Hence the exponential term of the matrix -i $\mathcal{H}_m$ L can be expended as
	\begin{align}
		e^{-i \mathcal{H}_m L} = \phi e^{-iLT} = \phi \left[a_0 \textbf{I} + a_1 (-iLT) + a_2 (-iLT)^2 \right] \nonumber \\
		= \phi \left[a_0 \textbf{I} - iLTa_1 - L^2T^2a_2 \right],
	\end{align}
	Here $T$ is the traceless matrix calculated from the Hamiltonian as $T = \mathcal{H}_m - {\rm Tr}(\mathcal{H}_m) {\bf I}/3$,  where ${\rm Tr} (\mathcal{H}_m) = E_{\nu} + A (1 + \epsilon_{ee} + \epsilon_{\mu\mu} + \epsilon_{\tau\tau})$ and $E_{\nu} = E_1 + E_2 + E_3$. The coefficients $a_{0,1,2}$ can be calculated in terms of eigenvalues of $T$ matrix, {\it i.e.}, $\lambda_a$, $a = 1,2,3$. One can finally write the evolution operator $\mathcal{U}_f$ in flavour state basis as
	
	\begin{widetext}
	\begin{equation}
		\mathcal{U}_m(L) = e^{-i \mathcal{H}_mL} = \phi \sum_{a=1}^{3}e^{-iL\lambda_a}\frac{1}{3\lambda_a^2 + c_1}\big[(\lambda_a^2 + c_1)\textbf{I} + \lambda_a T + T^2\big],
	\end{equation}
	
	\begin{equation}
		\mathcal{U}_f(L) = e^{-i \mathcal{H}_fL} = U_{\nu}e^{-i \mathcal{H}_mL}U_{\nu}^{-1} = \phi  \sum_{a=1}^{3}e^{-iL\lambda_a}\frac{1}{3\lambda_a^2 + c_1}\big[(\lambda_a^2 + c_1)\textbf{I} + \lambda_a \tilde{T} + \tilde{T}^2\big],
		\label{Uf}
	\end{equation}
	\end{widetext}
	where $\phi = e^{-i L {\rm Tr}(\mathcal{H}_m/3)}$, $\tilde{T} \equiv (U_{\nu}T U_{\nu}^{-1})$ and the coefficient $c_1 = {\rm det}(T) ~{\rm Tr}(T^{-1})$.

	\section{NSI effects on quantum coherence}\label{MI}
	The parameter $\chi$ defined in Eq. (\ref{coherence}) is probably the most convenient coherence-measure for neutrino experiments. In case of three-flavour neutrino oscillations, $d = 3$, {\it i.e.}, the maximal value of $\chi$ is 2. $\rho$ can be calculated using the neutrino state represented by $\ket{\psi (t)} \equiv \ket{\nu_{\alpha}(t)} = \sum_{i=1,2,3} \mathcal{U}_{fij}(t) \ket{\nu_{\beta}}$, with $j=1,2,3$ and $\beta=e,\mu,\tau$. Here $\mathcal{U}_{fij}$ are elements of the evolution operator given in Eq. (\ref{Uf}). For SM interactions, coherence can then be obtained as
	\begin{equation*}
		\chi^{SM} = \lim\limits_{\epsilon_{\alpha\beta}\to 0} \chi^{NSI}.
	\end{equation*}
	Also, the $l_1$-norm based coherence quantifier can be expressed in terms of observable neutrino survival and transition probabilities \cite{Song}. For example, if neutrino is produced initially in $\ket{\nu_{\mu}}$ state, then
	\begin{equation}
		\chi = 2 \big[\sqrt{P_{\mu e}(t)P_{\mu\mu}(t)} + \sqrt{P_{\mu e}(t)P_{\mu \tau}(t)} + \sqrt{P_{\mu\mu}(t)P_{\mu\tau}(t)}\big]
		\label{ChiProb}
	\end{equation} 
	with normalization constraint on probabilities, {\it i.e.}, $\sum_{\alpha} P_{\alpha\beta} = 1 = \sum_{\beta} P_{\alpha\beta}$ with $\alpha, \beta = e, \mu, \tau$. From above Eq. (\ref{ChiProb}), it can be seen that $\chi$ achieves its maximal value 2 only when all the flavours of neutrinos are equally probable to appear, {\it i.e.}, when $P_{\mu\mu} = P_{\mu e} = P_{\mu\tau} = 1/3$. If neutrino is found to have unit probability of being in a specific flavour, in that case $\chi$ becomes minimum or zero which means that the coherence will be lost completely. Therefore, in case of three-flavour neutrino oscillations the $\chi$ parameter is bounded as $0\leq \chi \leq 2$.
	
	\begin{figure*}[t]
		\centering
		{(a)\includegraphics[scale=0.5]{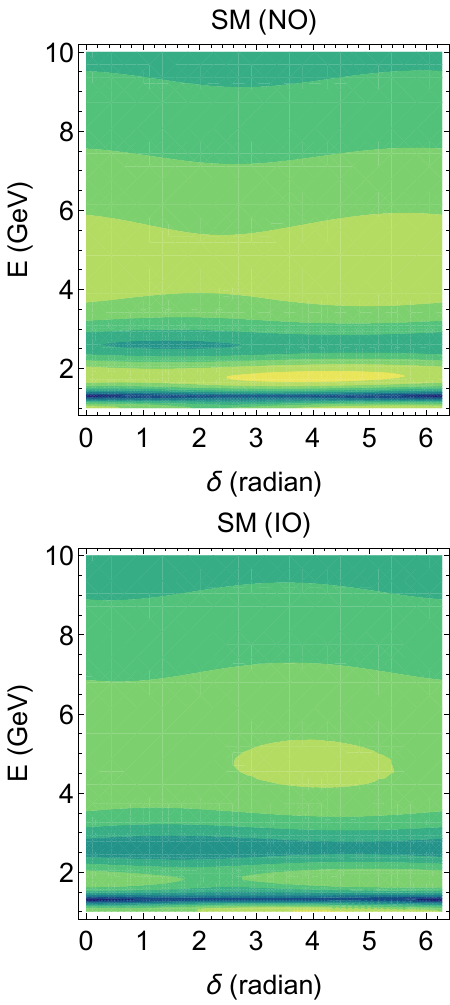} 
			(b)\includegraphics[scale=0.48]{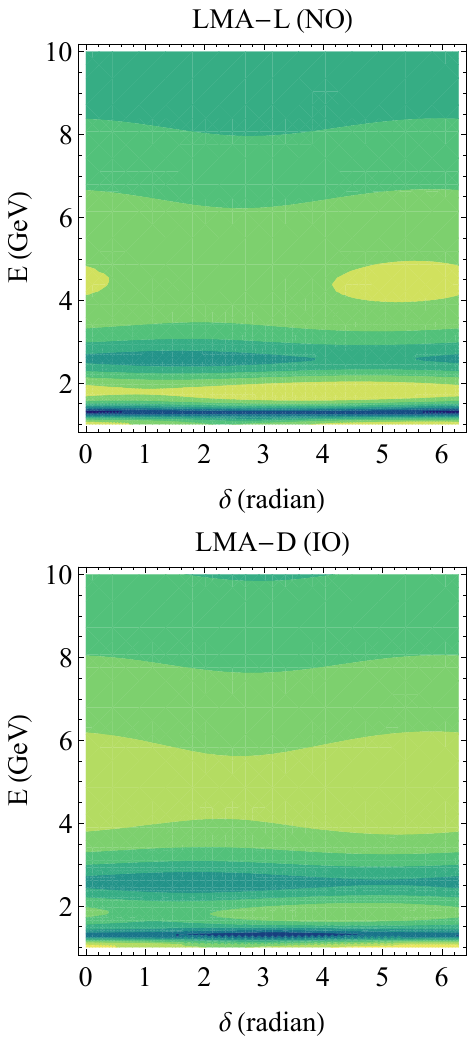}
			\includegraphics[width=15mm]{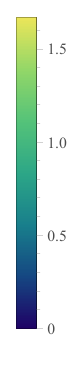}}
		\caption{The parameter $\chi$ plotted with respect to $E$ and $\delta$ in the context of DUNE ($L=1300 $ km \& $E=1-10$ GeV) experiment. (a) The upper and lower panels correspond to normal and inverted mass ordering, respectively representing the effect of SM interaction. (b) The upper and lower panels correspond to the LMA-Light+NO and LMA-Dark+IO combinations, respectively. The $\epsilon_{\alpha\beta}$ parameters are taken from recently updated analysis given in ref. \cite{Esteban3}. Minimum value (zero) of $\chi$ parameter represents the complete loss of coherence whereas the maximally coherent state is represented by $\chi = 2$.}
		\label{NSI}
	\end{figure*}
	
	In the following, we present our results for $\chi^{SM}$ and $\chi^{NSI}$ in the context of experimental set-up for upcoming long-baseline accelerator experiment DUNE (baseline $L = 1300$ Km). Hence we have $\alpha = \mu$ for accelerator $\nu_{\mu}$ beam and matter density potential is taken to be $A = 1.01 \times 10^{-13}$ eV ($\sim 2.8\,\rm g/cm^3$)\footnote{We converted the unit of matter density potential from $\rm g/cm^3$ to eV using $A = 7.6 \times \,Y_e\, \rho \times 10^{-14} $, where $Y_e = N_e/(N_e + N_n)$. Here $N_e$ and $N_n$ are the number densities of electrons and neutrons in the Earth and $\rho$ is the matter density in $\rm g/cm^3$. We  then have $Y_e \approx 0.48$ and $\rho = 2.8\,\rm g/cm^3$ which are convenient for Earth's matter as shown by Dziewonski et al. \cite{Dziewonski}. Hence, the value of $A$ turns out to be $ \approx 1.01 \times 10^{-13}$ eV.}. Further, oscillation parameters are as $\theta_{12} = 33.82^o$ (in case of SM interaction as well as for LMA-Light solution), $\theta_{23} = 49.6^o$, $\theta_{13} = 8.61^o$, $\Delta_{21} = 7.39 \times 10^{-5} \rm{eV}^2$ and $|\Delta_{32}| = 2.525 \times 10^{-3} \rm{eV}^2$ \cite{Esteban2}. Due to $CPT$-transformation given in Eq. (\ref{CPTvac}), the mixing-angle $\theta_{12}$ obtains the value $56.18^o$ for LMA-Dark solution.

	In Ref. \cite{Esteban}, the bounds on NSI parameters were obtained mainly by using constraints from observables such as the disappearance data from solar and KamLAND experiments, atmospheric neutrino data from Super-K, DeepCore and IceCube experiments along with the long-baseline (LBL) experimental data such as $\nu_{\mu}$ and $\bar{\nu}_{\mu}$ disappearance as well as $\nu_e$ and $\bar{\nu}_e$ appearance data from MINOS, $\nu_{\mu}$ and $\bar{\nu}_{\mu}$ disappearance data from T2K, $\nu_{\mu}$ disappearance data from NO$\nu$A experiment and also the data from COHERENT experiment. These observables are not sensitive to $\delta$-value and the sign of mass squared difference $\Delta_{31} (=m_3^2-m_1^2)$ and hence the NSI parameters were the same for both signs of $\Delta_{31}$. This analysis was updated in Ref. \cite{Esteban3} by including all relevant data in the neutrino sector which includes observables having functional dependence on the $CP$-violating phase as well as the sign of the  $\Delta_{31}$, {\it i.e.}, $\nu_e$ and $\bar{\nu}_e$ appearance data from T2K and NO$\nu$A. In this work, the allowed parameter space for NSI couplings was obtained for the light ($\theta_{12} < 45^o$) and dark ($\theta_{12} > 45^o$) octant. Following are the values of NSI parameters obtained for two favoured solutions \cite{Esteban3}: 
	\begin{itemize}
		\item LMA-Light sector with normal ordering: $\tilde{\epsilon}_{ee} = \epsilon_{ee} - \epsilon_{\mu \mu} \approx -0.1$, $\tilde{\epsilon}_{\tau\tau} = \epsilon_{\tau \tau} - \epsilon_{\mu \mu} \approx 0.01$, $\epsilon_{e\mu} \approx -0.06$, $\epsilon_{e\tau} \approx -0.1$, $\epsilon_{\mu\tau} \approx -0.01$.
		\item LMA-Dark sector with inverted ordering: $\tilde{\epsilon}_{ee} \approx -1.8$, $\tilde{\epsilon}_{\tau\tau} \approx -0.01$, $\epsilon_{e\mu} \approx 0.06$, $\epsilon_{e\tau} \approx -0.07$, $\epsilon_{\mu\tau} \approx -0.01$.
	\end{itemize}

	We study the impact of NSI on coherence parameter in view of the updated results obtained in \cite{Esteban3}. The results of our analysis are presented in Fig. \ref{NSI}. In this figure the observable quantifying coherence, $\chi$, is shown in the plane of the neutrino-energy $E$ (in GeV) and the $CP$-violating phase $\delta$ (in radian) for both positive (upper panel) and negative (lower panel) signs of $\Delta_{31}$. The range of $E$ (1 GeV $\leq E \leq 10$ GeV) along with the baseline length $L = 1300$ km correspond to the DUNE experimental set-up. 
	
	\begin{figure*}[t]
		\centering
		\includegraphics[scale=0.44]{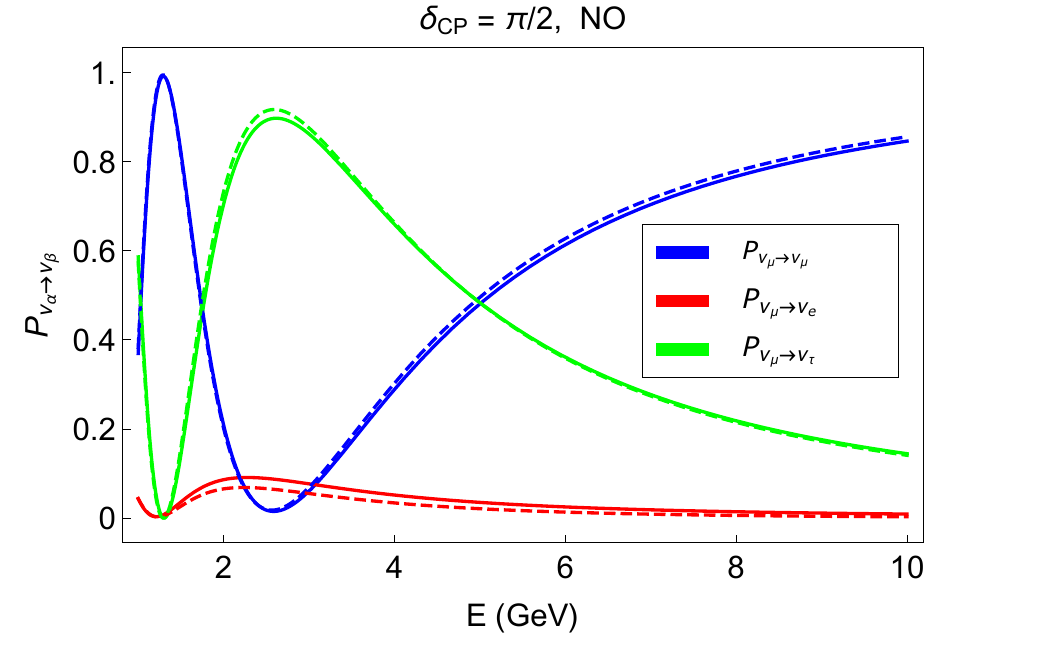} 
		\includegraphics[scale=0.44]{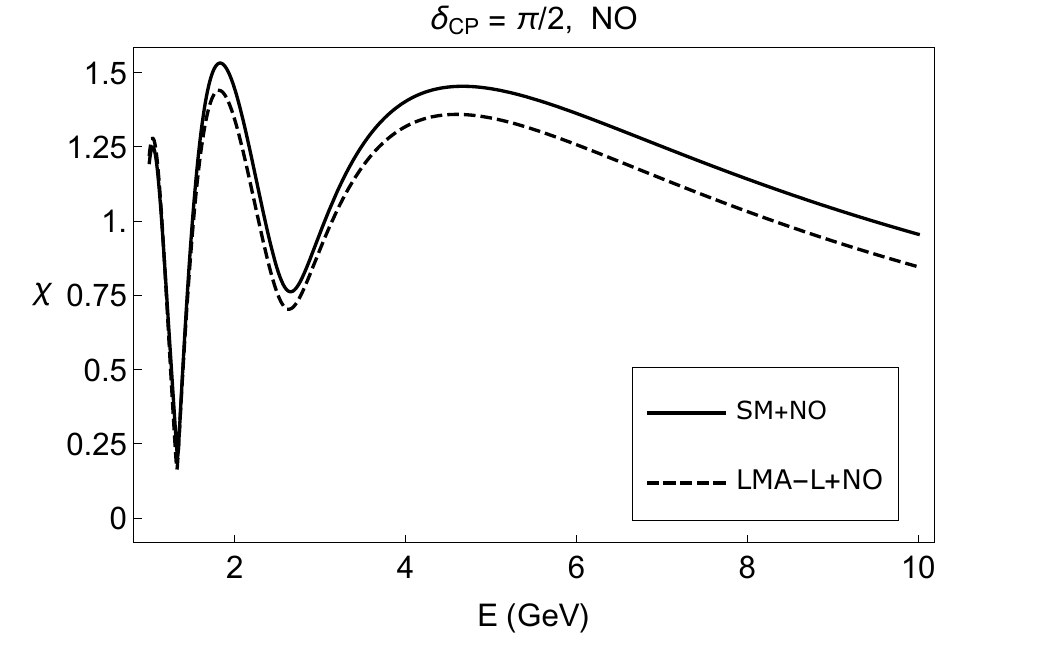}\\
		\includegraphics[scale=0.44]{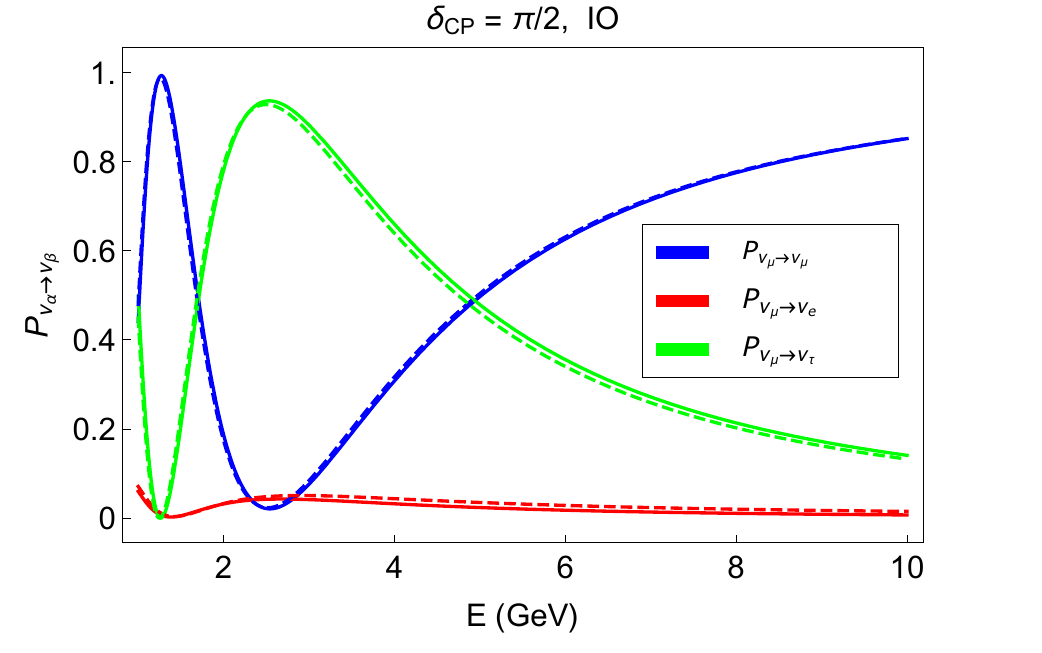} 
		\includegraphics[scale=0.44]{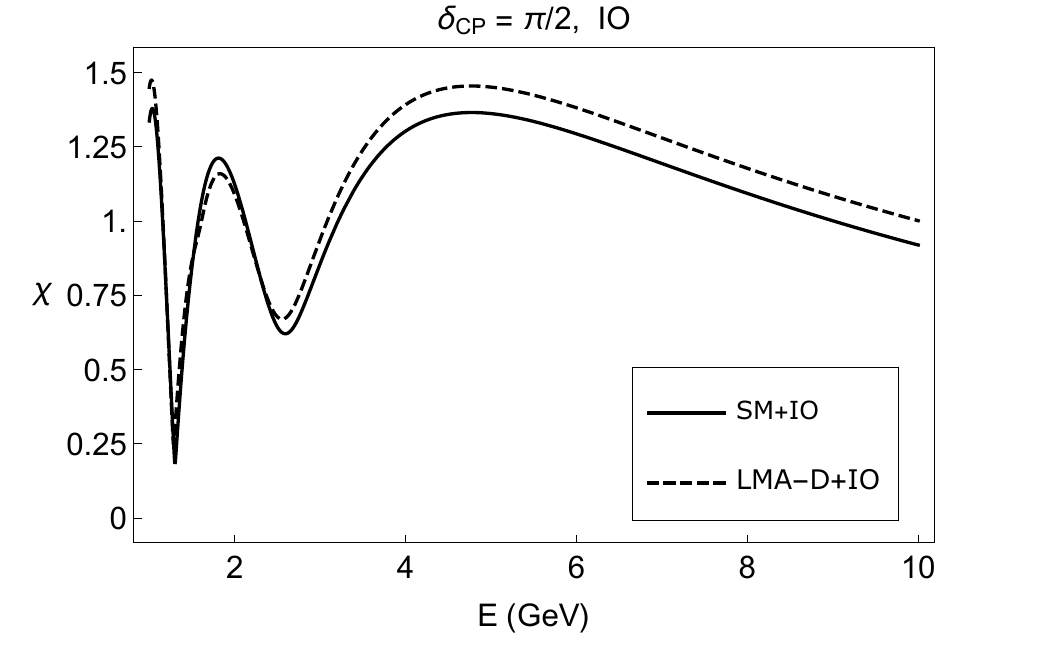}
		\caption{In the left panel probabilities $P_{\nu_{\mu} \rightarrow \nu_{\mu}}$ (blue), $P_{\nu_{\mu} \rightarrow \nu_e}$ (red) and $P_{\nu_{\mu} \rightarrow \nu_{\tau}}$ (green) are plotted with respect to $E$ in the context of DUNE ($L=1300 $ km) experiment for $\delta = \pi/2$  where solid and dashed lines correspond to the SM and NSI interaction, respectively. The upper and lower panels correspond to normal mass hierarchy (with SM and LMA-L scenario) and inverted mass ordering (with SM and LMA-D scenario), respectively. The right panel shows the variations of $\chi$ parameter with $E$ for $\delta = \pi/2$. The upper right and lower right panels correspond to the NO (with SM and LMA-L scenario) and IO (with SM and LMA-D scenario), respectively. The $\epsilon_{\alpha\beta}$ parameters are taken from recently updated analysis given in ref. \cite{Esteban3}.}
		\label{Prob}
	\end{figure*}
	
	The results shown in Fig. \ref{NSI}(a) correspond to the variation of coherence parameter for the SM interaction. It is observed that for the SM + NO, the maximal coherence achieved is $\chi_{max} \approx 1.67$ at around $E \sim 2$ GeV for $2.5 \lesssim \delta \lesssim 5.5$ which is approximately 84\% of the maximum allowed value of coherence, {\it i.e.,} 2.  Further, in the energy range 4 - 6 GeV (where the maximum neutrino flux is expected for the DUNE experiment), $\chi^{SM}$ can achieve quite large value ($\approx$ 1.5) for all values of $CP$-violating phase. For the SM + IO scenario, the maximal value of coherence is 1.5 in the energy range  4 - 5 GeV for $3 \lesssim \delta \lesssim 5.5$. This is 75\% of the maximum allowed value of the coherence parameter. Thus we see that within  SM, the quantumness of system which we have quantified in terms of coherence, is sensitive to the sign of $\Delta_{31}$ as well as the $CP$ violating phase and the system will have relatively large coherence (or quantumness) in case of NO.

	The upper plot in Fig. \ref{NSI}(b) depicts $\chi^{NSI}$ for LMA-Light solution with NO. It can be seen that the overall effect of this NSI solution is to reduce the coherence. The maximal coherence obtained for the LMA-L + NO solution is $\approx 1.58$ (79\% of the maximum allowed value) at around $E \approx 2$ GeV (for all $\delta$) and $E = 4 - 5$ GeV (for $0 \leq \delta \leq 0.5$ and $4 \leq \delta \leq 6.28$). This means that $\chi_{max}$ for the LMA-L + NO scenario is reduced by 5\% as compared to the SM + NO. Contrarily, for IO, the LMA-D solution (lower panel of Fig. \ref{NSI}(b)) provides slightly larger value of $\chi_{max}$ for all values of $\delta$ in the energy range 4 - 6 GeV in comparison to the SM. Moreover, in this energy range, the overall enhancement in the coherence of the system for the LMA-D + IO solution is around 6\% as compared to the SM + IO for all values of $\delta$ except for $2.5 \lesssim \delta \lesssim 5.5$.

	In the left panel of Fig. \ref{Prob} we have plotted the survival probability $P_{\nu_{\mu} \rightarrow \nu_{\mu}}$ (blue) and transition probabilities $P_{\nu_{\mu} \rightarrow \nu_e}$ (red) and $P_{\nu_{\mu} \rightarrow \nu_{\tau}}$ (green) with respect to $E$ with $L = 1800$ km for DUNE. The upper and lower panel of Fig. \ref{Prob} represent the case of NO (with SM (solid lines) and LMA-L (dashed lines)) and IO (with SM (solid lines) and LMA-D (dashed lines)), respectively. The right panel in Fig. \ref{Prob} depicts the variation of $\chi$ parameter with respect to the neutrino-energy $E$ ($L = 1800$ km). 
		We have taken the value of CP-violating phase $\delta$ to be $\pi/2$. It is evident from the figure that the deviation in the coherence parameter $\chi$ from its SM value due to NSI effects is larger in comparison to the probabilities. This is true for both NO and IO scenarios. For example, the maximum difference observed in $P_{\nu_{\mu} \rightarrow \nu_{\mu}}$ is around 2\%, while this difference is higher (around 5\%) in case of $\chi$ parameter for NO. Similarly, for IO, the difference observed in $P_{\nu_{\mu} \rightarrow \nu_{\mu}}$ is less then 2\%, while the coherence is increased by $\sim 6\%$ for the LMA-D scenario as compared to the SM. Therefore a small deviation in probability due to NSI effects can trigger a larger deviation in the coherence parameter. Hence, it is worth re-examining the NP effects on various measurements of quantum correlations. In particular, the nature of correlations viz. entanglement and non-locality in terms of Bell-type and Leggett-Garg inequalities in neutrino oscillations should be reanalyzed under NSI effects.  To test these correlations, the long baseline (LBL) experiments can serve the purpose as it is possible to have good control over the neutrino-source and the detector. Among numerous LBL facilities, DUNE will be more sensitive to NSI effects since neutrinos will have to travel longer distance in Earth's material medium. Thus the  NSI effects will be more notable in the correlation measures for the DUNE experimental setup.

	\section{Conclusions}\label{Conclusion}
In this work we study the impact of new physics on quantum coherence in the system of oscillating neutrinos. Quantum coherence quantifies the quantumness embedded  in the system and is considered to be a resource theoretical tool which can be utilized in quantum algorithms and quantum channel discrimination. Further, it is intrinsically related to several witnesses of non-classicality. The NSI effects on coherence are incorporated in a model-independent way using the formalism of effective Lagrangian where higher dimensional operators have been added. 
Recently, a global analysis of all relevant neutrino oscillation data in the presence of NSI was performed in \cite{Esteban3}. This analysis included constraints from $\nu_e$ and $\bar{\nu}_e$ appearance data from T2K and NO$\nu$A due to which the allowed NSI parameter space is now different for normal and inverted mass orderings.
Further, it was observed that two scenarios, LMA-Light solution with normal ordering and LMA-Dark solution  with inverted ordering, provide a good fit to all data. Using these inputs, we analyse new physics effects  in the context of DUNE experimental setup ($L = 1300$ km \& $E = 1 - 10$ GeV).  Specific results include the following:
\begin{itemize}
	\item The LMA-Light solution with normal ordering decreases the value of coherence parameter in comparison to the SM with normal ordering.
	
	\item For the LMA-Dark solution with inverted ordering, the coherence in the system is enhanced in comparison to the case of SM with inverted ordering for $E \approx 4$ GeV, the energy corresponding to maximum neutrino flux at DUNE, for almost all values of $CP$ violating phase.
	
	\item Neutrinos can maintain large coherence over macroscopic distances in case of both SM and NSI. However, the oscillating system of three flavours of neutrinos would never reach its maximum value $\chi = 2$ in the context of DUNE experiment.
	
	\item A small change in probabilities due to NSI effects can trigger relatively large alteration in the coherence inherent in the neutrino-system. 
	
\end{itemize}
In summary, we find that for inverted ordering, NSI effects can facilitate quantum information tasks owing to increase in coherence whereas for normal ordering, new physics effects can decrease the coherence. Moreover, as quantum coherence is related to various measures of quantum correlations viz. entanglement and non-locality, it would be worth re-examining studies related to these observables in the presence of new physics as a small change in neutrino transition probabilities can greatly revise the coherence, in other words the quantumness of the system. The methodology given in this work can be implemented for such studies.

	\section*{Acknowledgments}
	We are thankful to the organizing committee of the \textit{IITB - ICTP Workshop on Neutrino Physics -2018} where initial ideas related to this work were generated. We would also like to thank S. Uma Sankar for providing useful insights into the non-standard interactions in neutrinos.

\end{document}